\documentclass[amsmath,amssymb,floatfix,prb,epsf,twocolumn,superscriptaddress]{revtex4}
\usepackage{amsmath,amssymb,natbib,bm,graphicx,url,epsfig,wasysym}
\usepackage[ansinew]{inputenc}
\usepackage{bm}
\usepackage{color}
\newcommand{\be}{\begin{equation}}
\newcommand{\ee}{\end{equation}}
\newcommand{\bes}{\begin{equation}\begin{split}}
\newcommand{\ees}{\end{split}\end{equation}}
\newcommand{\bea}{\begin{eqnarray}}
\newcommand{\eea}{\end{eqnarray}}
\newcommand{\nn}{\nonumber}

\DeclareMathOperator{\curl}{curl}

\def\beq{\begin{equation}}
\def\eeq{\end{equation}}
\def\bea{\begin{eqnarray}}
\def\eea{\end{eqnarray}}
\def\ena{\end{eqnarray}}

\begin{document}


\title{ Statistical transmutation in Floquet  driven optical lattices}

\author{Tigran Sedrakyan}
\affiliation{William I. Fine Theoretical Physics Institute,  University of Minnesota, Minneapolis, Minnesota 55455, USA}
\affiliation{Physics Frontier Center and Joint Quantum Institute, University of Maryland, College Park, Maryland 20742, USA}

\author{Victor Galitski}

\affiliation{ Joint Quantum Institute and Condensed Matter Theory Center, Department of Physics,
University of Maryland, College Park, Maryland 20742-4111, USA}

\affiliation{School of Physics, Monash University, Melbourne, Victoria 3800, Australia}

\author{Alex Kamenev}
\affiliation{William I. Fine Theoretical Physics Institute,  University of Minnesota, Minneapolis, Minnesota 55455, USA}

\begin{abstract}
We show that interacting bosons in a periodically-driven two dimensional (2D)
optical lattice may effectively exhibit fermionic statistics. The phenomenon is
similar to the celebrated Tonks-Girardeau regime in 1D. The Floquet band of a
driven lattice develops the moat shape, i.e. a minimum along a closed contour in
the Brillouin zone. Such degeneracy of the kinetic energy favors fermionic
quasiparticles. The statistical transmutation is achieved by the Chern-Simons
flux attachment similar to the fractional quantum Hall case. We show
that the velocity
distribution of the released bosons is a sensitive probe of the fermionic
nature of their stationary Floquet state.

\end{abstract}

\date{\today}

\maketitle


It is  well established  \cite{Tonks,Girardeau,Lieb,CNY,Gaudin,exp1,exp2} that a one dimensional (1D) Bose gas 
with a strong short range repulsion (or equivalently small density) - the Tonks-Girardeau gas, exhibits features of weakly interacting
fermions. Can a similar phenomenon take place and be observed in dimensions larger than one? 
One example of statistical transmutation in 2D is provided by the composite boson picture of the 
fractional quantum Hall effect (FQHE) \cite{AMP,Jain,lopez,composite,Shankar,Simon}. In this case the kinetic energy is totally degenerate due to the Landau quantization and the ground state is solely determined by the interactions. It was recently shown\cite{SKG,SGK1,chiral} that it is actually enough to have the kinetic energy degenerate along a {\em line} in the 2D reciprocal space (so called ``moat'') to achieve the statistical transmutation\cite{AMP}. The moat-like dispersion was discussed earlier in context  of  Rashba spin-orbit coupling\cite{SKG, Rashba,GS,Ian1,Berg}, and for certain lattices with more than one site per unit cell and next nearest neighbors hopping\cite{SGK1,chiral,varney}. 

In this paper we show that the moat dispersion may be found even in simplest lattices (e.g. square) upon 
a suitable periodic driving. The lowest Bloch band adiabatically evolves into a Floquet band, which exhibits an approximately flat minimum along a closed contour encircling the $\Gamma$-point. Consequently the single particle density of states (DOS) diverges as inverse square root of energy at the bottom of the moat. Such 1D behavior of 2D DOS motivates the analogy with the Tonks-Girardeau case. The specific mechanism of the statistical transmutation in 2D, however, is very different from its 1D sibling. While the latter is achieved by simple sign inversion of the wave function at coinciding coordinates of particles, the former requires flux attachment, similar to FQHE. As a result, the emerging 2D fermions are subject to the time reversal symmetry breaking effective magnetic field, and leading to a peculiar Landau spectrum.  We show that both the fermionic nature of the state and the effective magnetic field may be detected through the velocity distribution of the released gas.

The Hamiltonian of interacting Bose gas with the moat dispersion relation can be written as 
\bea
\label{BH}
H=\frac{1}{2M}\sum_{\bf r}\left[\Phi_{\bf r}^\dagger\left(|\hat{\bf k}|-k_0\right)^2\Phi_{\bf r}
 +{2g}(\Phi_{\bf r}^\dagger\Phi_{\bf r})^2\right],
\ena
where $M$ is the mass of bosons, $k_0$ is the  radius of the degenerate minima of the dispersion, $g$ is the dimensionless contact interaction strength  
and $\Phi_{\bf r}^\dagger$, $\Phi_{\bf r}$ are correspondingly 
boson creation and annihilation operators. 



\begin{figure}[t]
\centerline{\includegraphics[width=90mm,angle=0,clip]{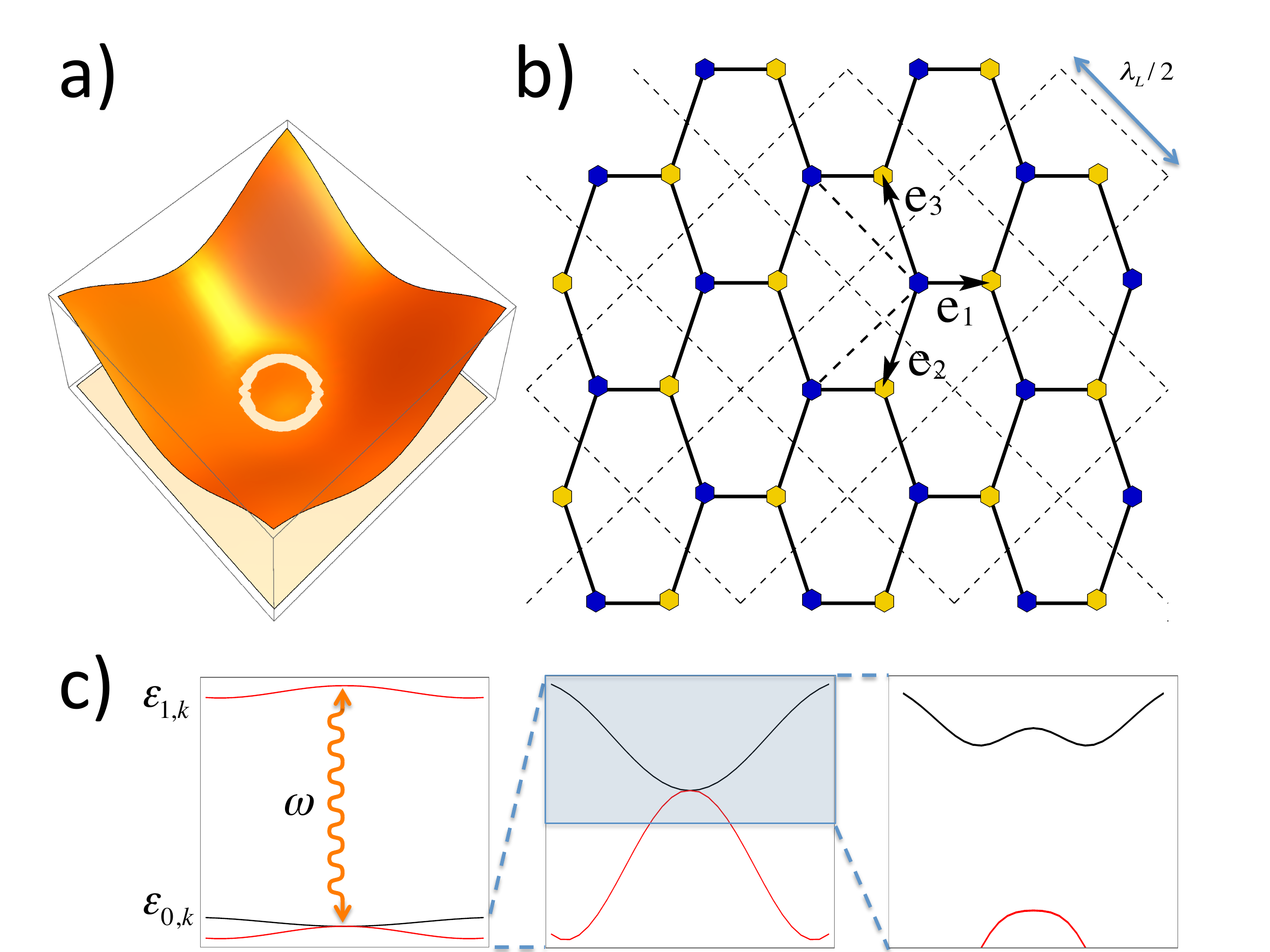}}
\caption{(Color online) 
a) Lowest Floquet band exhibiting approximately flat moat. 
b) Fragment of a square optical lattice with two sites per unit cell\cite{tarruell} that gives raise to an approximately flat moat
in the lowest Floquet band upon resonant driving.
c) Left panel shows the band structure of the un-driven optical lattice. 
The driving frequency $\omega$ is of order of the gap causing resonant coupling between the lowest and the first excited bands at ${\bf k}=0$.
Middle panel: enlargement of the lowest and shifted first excited bands. Right panel: Blow up of the driven lowest Floquet band exhibiting a double-well feature in $k_x+k_y$ 
direction indicating appearance of the moat in the Brillouin zone.
}
\label{moat}
\end{figure}


To engineer a system of spinless bosons effectively described by the  Hamiltonian~(\ref{BH}), 
we adopt the idea of Floquet  optical lattice shaking developed in  Ref.~\onlinecite{chin}. The authors 
employed harmonically shifted  1D optical lattice  at a near-resonance 
frequency corresponding to the lowest-band to the first-excited-band transition. As a result they have created a 1D dispersion relation with double minimum. The 2D generalization of this strategy is discussed below.  
A straightforward idea of using simple square lattice with the potential $\sim \sin^2(k_L x)+\sin^2(k_L y)$ does not work because of the separable (in $x,y$) nature of the potential. Indeed the corresponding bands are labeled by the two-integers $\epsilon_{n_x,n_y}$ with the lowest band being $\epsilon_{0,0}$. The low energy excited bands $ \epsilon_{0,1}$ are $\epsilon_{1,0}$ are highly anisotropic and thus do not lead to a flat moat. The next  excited band $\epsilon_{1,1}$ is approximately isotropic, but shaking the lattice either in $x$ or $y$ directions does not produce matrix elements between $\epsilon_{1,1}$ and $\epsilon_{0,0}$, due to orthogonality of the separable wave functions.

The simplest way to engineer an approximately flat moat is to create a square optical lattice with two sites per unit cell. Such a lattice can be constructed by 
fusing a laser setup with laser beams of wavelength $\lambda_L=2\pi/k_L$, resulting in a standing wave intensity pattern forming a regular square lattice, 
and then adding two additional identical laser beams directed along $x+y$ and $x-y$ diagonals (X scheme). 
If the lasers directed along $x$, $y$, and $x-y$ directions are calibrated to have exactly the same phases while the laser along $x+y$ diagonal has a 
phase shift amounting to $\pi$, 
such setup results in a potential
\bea
\label{pot}
U(x,y)\!&=&\! U_0 \Bigl[\sin^2(k_L x)+\sin^2(k_L y)\nn\\
&+& \!\sin^2[k_L (x-y)]+\cos^2[k_L (x+y)]\Bigr].
\eea
This potential realizes a square optical lattice of the depth $U_0$, having two sites per unit cell (A and B), 
with vectors $k_L{\bf e}_1=(\frac{\pi}{3},\frac{\pi}{3})$,  $k_L{\bf e}_2=(\frac{\pi}{3},-\frac{2 \pi}{3})$ 
and   $k_L{\bf e}_3=(-\frac{2 \pi}{3},\frac{ \pi}{3})$, depicted in Fig.~1. 
One can harmonically drive  this  lattice by  applying  time dependent phase shifts to the lasers, resulting into the transformation $x\rightarrow x-\Delta \cos\omega t$, $y\rightarrow y-\Delta \cos\omega t$, where $\Delta$ is the shaking amplitude.
Below we show that, if the frequency $\omega$ is nearly resonant with two lowest bands, 
an almost flat moat appears in the lowest Floquet band.

The Lagrangian describing such a time-dependent problem is given by 
\bea
\label{A11}
{\cal L}=\bar\psi({\bf r},t)\Big\{i \partial_t  -\hat{H}(t)\Big\}\psi({\bf r},t)\nn\\
\hat{H}(t) =-\frac{\partial_{\bf r}^2}{2 M}
+U\left[x-x(t),y-y(t)\right],
\eea
where  
$\psi({\bf r},t) $ is a single particle wave function. 
Periodicity of lattice 
with $T_r= \pi/k_L$, 
implies that the solution  $\psi_{n,{\bf k}}({\bf r},t)$  of the Shr\"{o}dinger equation 
 with Hamiltonian $\hat{H}(t)$, can be uniquely described by the space-time periodic single particle eigenstate
$\varphi_{n,{\bf k}}({\bf r},t)$
of the Floquet operator $i \partial_t-H(t)$ 
 as $\psi_{n,{\bf k}}({\bf r},t)=e^{i E_n({\bf k}) t}e^{i {\bf k}{\bf r}}\varphi_{n,{\bf k}}({\bf r},t)$.
This representation is analogous to the Bloch representation of states in time independent periodic potentials. 
Here $E_{n}({\bf k})\in(0,\omega)$ is the Floquet energy 
of $n$'th Bloch band,
defined modulus of multiples of $\omega$ (corresponding to the energy quanta produced by driving ).
The Bloch momentum is given by ${\bf k}=2\pi {\bf m}/L,\; {\bf m}=(m_x,m_y)$, $m_{x,y}=1,\ldots, L$, with $L$ being the lattice size. 
Here the function $\varphi_{n,{\bf k}}({\bf r},t) $ is periodic in $x$ and $y$ directions with the period $T_r$, and in time with the period $T=2\pi/\omega$.
To find the Floquet spectrum we expand the
periodic counterpart,  $\varphi_{n,{\bf k}}({\bf r},t)$, of the Bloch eigenfunction in Fourier series 
over momenta $\frac{2 \pi m_x}{T_r}$,  $\frac{2 \pi m_y}{T_r} $ 
 and energies $\frac{2 \pi s}{T} $ with integer $s, m_x, m_y\in(-\infty, \infty)$ and numerically diagonalize the Hamiltonian in this basis, assuming it is a large finite dimensional matrix.

%


To understand the origin of the moat dispersion qualitatively consider the two lowest bands,  $\epsilon_{0}({\bf k})$ 
and $\epsilon_{1}({\bf k})$, of the potential (\ref{A11}).  Their Bloch functions $\varphi_{0,{\bf k}}({\bf r})$ and $\varphi_{1,{\bf k}}({\bf r})$ are correspondingly even and odd functions with respect to interchange of A and B sites of the unit cell, i.e. $x+y\to -x-y$. Figure~1$c$ depicts these two bands for $U_0=11 E_R$, where $E_R= k_L^2/2M$ is the recoil energy.
The resonant shaking of the phases with frequency  $\omega \approx \epsilon_{1}(0)-\epsilon_{0}(0)\approx 2.32 E_R$ shifts the upper band as $\epsilon_{1}({\bf k})-\omega$ to touch the lower one at ${\bf k}=0$ and induces the off diagonal matrix element $V_{\bf k}$ between these two bands. Due to the parity of the Bloch  functions,  the 
latter originates from the last term in Eq.~(\ref{pot}) and is given by 
$V({\bf k})=\frac{U_0}{2}J_1(4 k_L \Delta) \int d^2{\bf r}\, \bar\varphi_{0,{\bf k}}({\bf r}) \sin[2 k_L(x+y)]\varphi_{1,{\bf k}}({\bf r})$, where $J_1$ is the Bessel function. 
The effective two-band Hamiltonian 
\bea 
\label{A2}  
{\cal H}_{\bf k}=
\left( \begin{array}{cc}
            \epsilon_{0}({\bf k}) & V({\bf k}) \\
             \bar V({\bf k}) &\epsilon_{1}({\bf k})-\omega
            \end{array}
\right).
\ena
yields the moat shape for the band, which is adiabatically connected to the original lowest Bloch band  $\epsilon_{0}({\bf k})$. Figure~\ref{radius}a shows dependence of the characteristic radius of the moat as a function of  $U_0$ at two values of $k_L \Delta/\pi=0.01$ and $0.02$. The moat 
appears at finite values of $U_0/E_R$ and exists until about $U_0/E_R \lesssim 14$. Figure \ref{radius}b  shows the the ratio of the difference of maximal and minimal energies along the moat, $E_{\text{max}}-E_{\text{min}}$, and the hight of spectrum in the $\Gamma$ point, $E_{\Gamma}-E_{\text{min}}$, illustrating the flatness of the moat as function of $U_0$.


An exactly flat Floquet moat band may be achieved by rotating of the lasers\cite{so} in 1D optical lattices 
with the same resonant frequency as specified in Ref.~[\onlinecite{chin}]. 




\begin{figure}[t]
\centerline{\includegraphics[width=90mm,angle=0,clip]{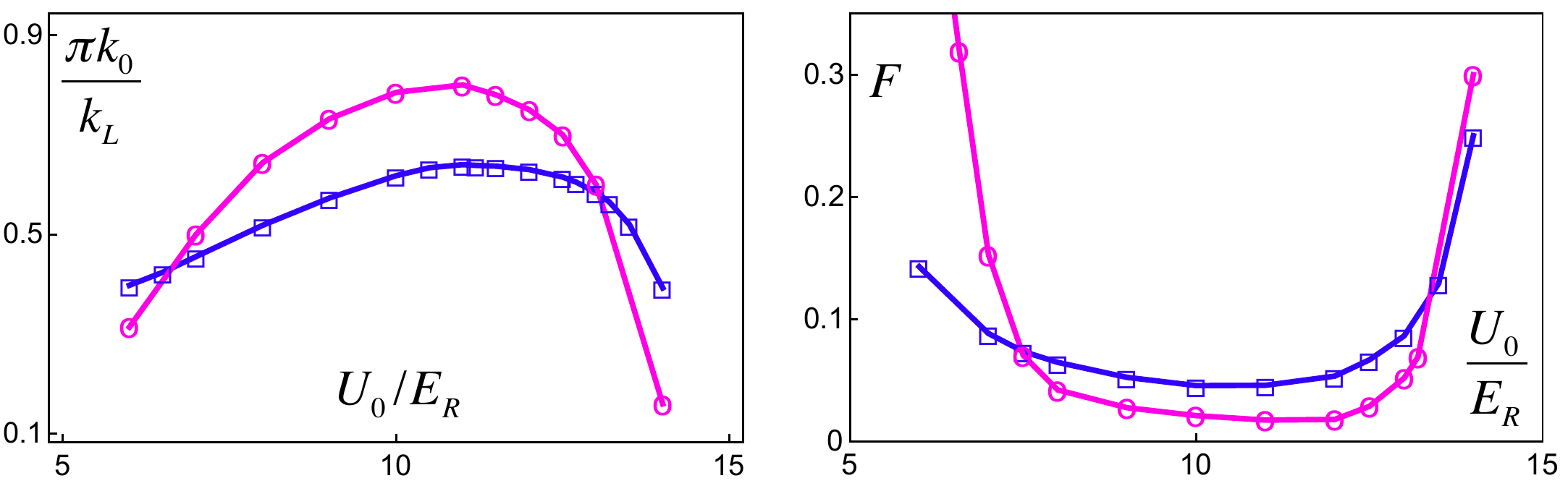}}
\caption{(Color online) Left panel: characteristic radius $k_0$ plotted as a function of  dimensionless $U_0/E_R$ at two values of 
$k_L \Delta/\pi=0.01$ (circles) and $0.02$ (squares). Right panel:  the flatness measure $F$ of the moat versus $U_0/E_R$ is shown. 
For definition see the main text.}
\label{radius}
\end{figure}



Assuming an ideal moat, the interacting bosons are described by the Hamiltonian (\ref{BH}). Its single particle
density of states diverges as $\epsilon^{-1/2}$, similar to the 1D case. This immediately suggests that the condensate can not be stable at any finite temperature. A more rigorous consideration, Ref.~[\onlinecite{Brazovskii}], starts from assuming a condensate in a state ${\bf k}_0$ along the moat and derives the anisotropic spectrum of Bogoliubov excitations $E_{\bf p}=\left[{\epsilon_{\bf p}^2 + 2 (g n/M) \epsilon_{\bf p}}\right]^{1/2}$, 
where  $\epsilon_{\bf p} = (|{\bf p}+{\bf k}_0|-k_0)^2/2M $. It is easy to see then that the fraction of thermally excited quasiparticles diverges as a power law (Ref.~[\onlinecite{Brazovskii}] deals with 3D, where divergence is only logarithmic) with the system size, indicating instability of the condensate at any finite temperature.  

Moreover, there are compelling reasons to believe (see also Ref.~\onlinecite{Lamacraft}) that the groundstate 
of the repulsively interacting bosons with the moat dispersion does not break $U(1)$ symmetry but breaks the time reversal symmetry. Thus the state is not 
represented by the condensate. It was argued in Refs.~[\onlinecite{SKG,SGK1,chiral}] that the better approximation to the actual ground state is the variational wave function, given by a Landau level fully filled with fermions, which are transformed to bosons with the Chern-Simons (CS) flux attachment:
\bea
\label{ca0}
\Phi({\bf r}_1,\cdots {\bf r}_N)=e^{i \sum_{i<j}\arg[{\bf r}_i-{\bf r}_j]}\Psi_F({\bf r}_1,\cdots {\bf r}_N).
\ena
Here $ \Psi_F({\bf r}_1,\cdots {\bf r}_N)$ is a fully antisymmetric state of $N$ fermions. By this reason the wave function (\ref{ca0}) does not cost {\em any} short-range interaction energy.  
The CS phase $e^{i \arg[{\bf r}_i-{\bf r}_j]}=(z_i-z_j)/|z_i-z_j|$,
with complex $z_j=x_j+i y_j$, is antisymmetric with respect to the exchange of any two coordinates, restoring the bosonic nature of  $\Phi({\bf r}_1,\cdots {\bf r}_N)$.  
The CS factor costs kinetic energy, since it effectively modifies the momentum operator in Eq.~(\ref{BH}) as   $i\nabla _{\bf r} \rightarrow  i\nabla _{\bf r}-{\bf A}({\bf r})$, where the vector potential is given by
\bea
\label{A}
{\bf A}_{\alpha}({\bf r}_i)=\varepsilon_{\alpha\beta}\sum_{j\neq i} \frac{({\bf r}_i-{\bf r}_j)_\beta}
{|{{\bf r}_i}-{{\bf r}_j}|^2},
\ena
and $\varepsilon_{\alpha\beta}$ is an antisymmetric tensor.  The CS magnetic field, originating from ${\bf A}$ is given by 
$B({\bf r}_j)=\curl {\bf A}({\bf r}_j)=2\pi\sum_{i\neq j}\delta({\bf r}_i-{\bf r}_j)\equiv2\pi n({\bf r}_j)$, where $n({\bf r})$ is the density operator.

\begin{figure}[t]
\centerline{\includegraphics[width=80mm,angle=0,clip]{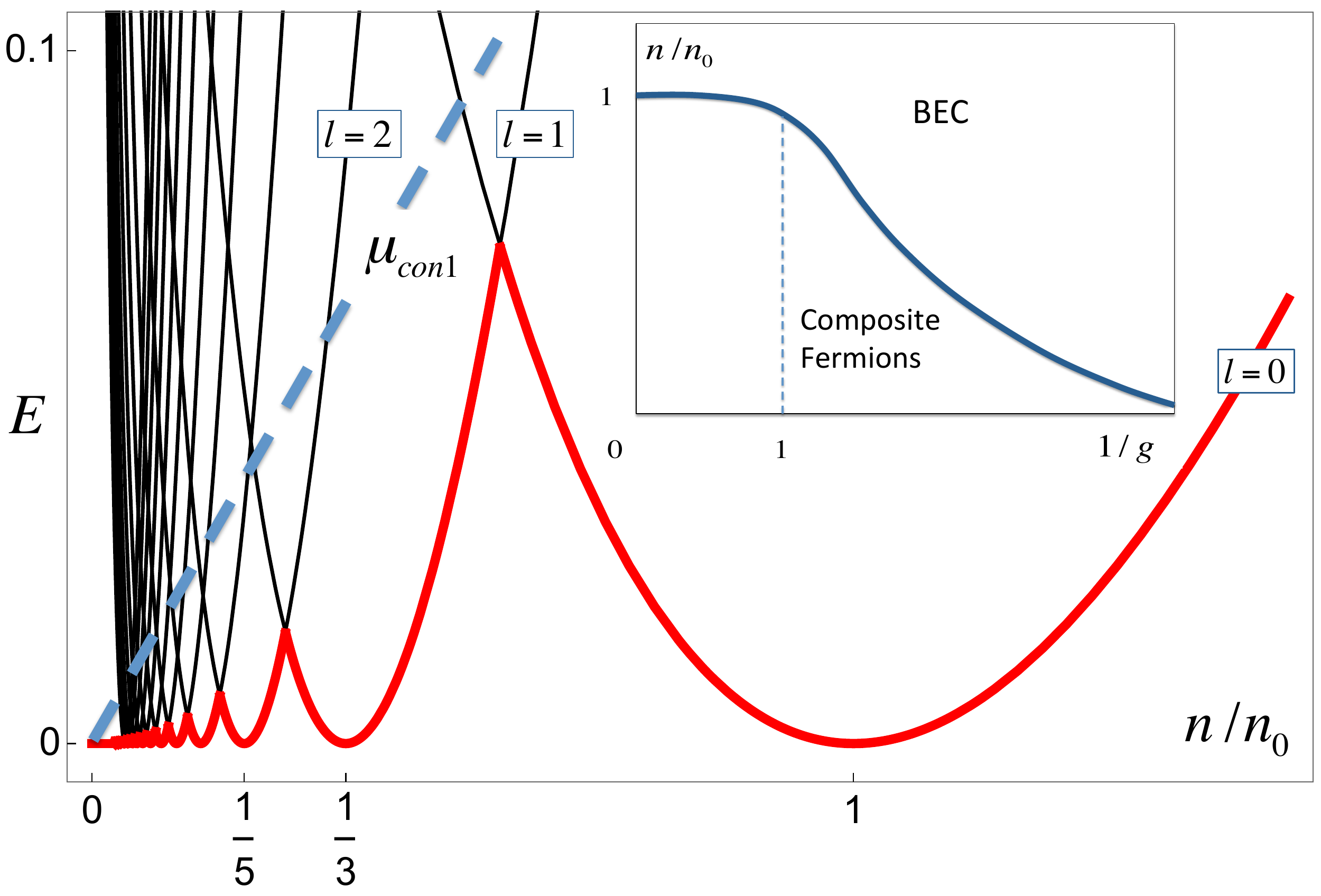}}
\caption{(Color online) Excitation spectrum in units of $E_R$ is plotted versus  $n/n_0$. 
The per particle energy, $E_{gs}/E_R$, represented by the fully filled lowest LL,  
is shown by the thick grey (red) curve. The thin black lines represent excited energies.
Dashed line represents the chemical potential of a condensate $\mu_{\text{con1}}/E_R$ corresponding to the interaction parameter $g\sim 1$. For such strong interactions the transition from composite fermion state to zero temperature BEC  takes place at densities $n/n_0\sim 1$. Upon lowering $g$, the transition shifts towards smaller $n/n_0$, as schematically shown in the inset.}
\label{fig3}
\end{figure}

By analogy with the  fractional quantum Hall effect we first adopt a mean field approximation, which  replaces a set of the flux lines with a {\em uniform} magnetic field  $B({\bf r})\rightarrow B=2\pi n$, where $n$ is the average density. In this approximation $\Psi_F$ is a state of non-interacting fermions with the dispersion relation $(|{\bf k}|-k_0)^2/2M$ placed in a uniform magnetic field $B$. The energies of corresponding Landau levels are given by ${E_l}=\left(k_0^2/2M\right)\left[\sqrt{\frac{\omega_c}{\left(k_0^2/2M\right)}(l+1/2)}-1\right]^2$, where 
 $\omega_c=B/M=2 \pi n/M$ and and $l=0,1,2,\ldots$.  The corresponding spectrum, shown in Fig.~\ref{fig3}, consists of set of non-monotonic functions of the field (i.e. density), which reach zero energy at 
\bea
\label{fil}
n_l=\frac{k_0^2}{4 \pi (l+1/2)}.
\ena
At this particular set of densities the fermionic wave function is given by the fully occupied (indeed there is exactly 
one flux quanta per particle)  $l$-th Landau level:    
\bea
\label{LL}
\Psi_F^{(l)}({\bf r}_1,\cdots {\bf r}_N)=\frac{1}{\sqrt{N!}} \det_{m,j} \Big[\chi_{m}^{(l)}({\bf r}_j)\Big], 
\ena
where $\chi_{m}^{(l)}({\bf r})= ({2\pi(l+m)! l!})^{-1/2} (b^\dagger)^{l+m} (a^\dagger)^l\big[e^{-\frac{1}{4}|z|^2}\big]$, is a state with the angular momentum  $m= -l \cdots -l+N$ at the Landau level $l$. Here 
$a^\dagger=\frac{1}{\sqrt{2}}(\frac{\bar{z}}{2}-2 \frac{\partial}{\partial z})$, 
$b^\dagger=\frac{1}{\sqrt{2}}(\frac{z}{2}-2 \frac{\partial}{\partial\bar{z}})$ are the ladder operators.  
For intermediate densities
$n_{l+1} < n < n_{l}$
the ground state is a mixture of two fermion liquids with  densities   $n_{l}$ and $n_{l+1}$. The variational state (\ref{ca0}), (\ref{LL}) breaks time-reversal symmetry due to the effective Chern-Simons magnetic field, however it conserves the $U(1)$.  
Though within the mean-field approximation the state has zero energy,  its  actual energy is given by the expectation value of the Hamiltonian operator (\ref{BH}) (or rather only its kinetic energy part). The corresponding calculations, given in the supplementary material, show that the energy  per particle scales as $E(n)\sim \frac{k_0^2}{M}\, \left(\frac{n}{n_0}\right)^2\log^2({n}/{n_0})$.  This should be compared with either the naive
estimate for the condensate state $E\sim gn/M$, or the result of Ref.~[\onlinecite{Lamacraft}] $E\sim n^{4/3}$. In any 
event, the composite fermion variational state  (\ref{ca0}), (\ref{LL})  is seen to be advantageous at small enough density.\cite{jura}



\begin{figure}[t]
\centerline{\includegraphics[width=85mm,angle=0,clip]{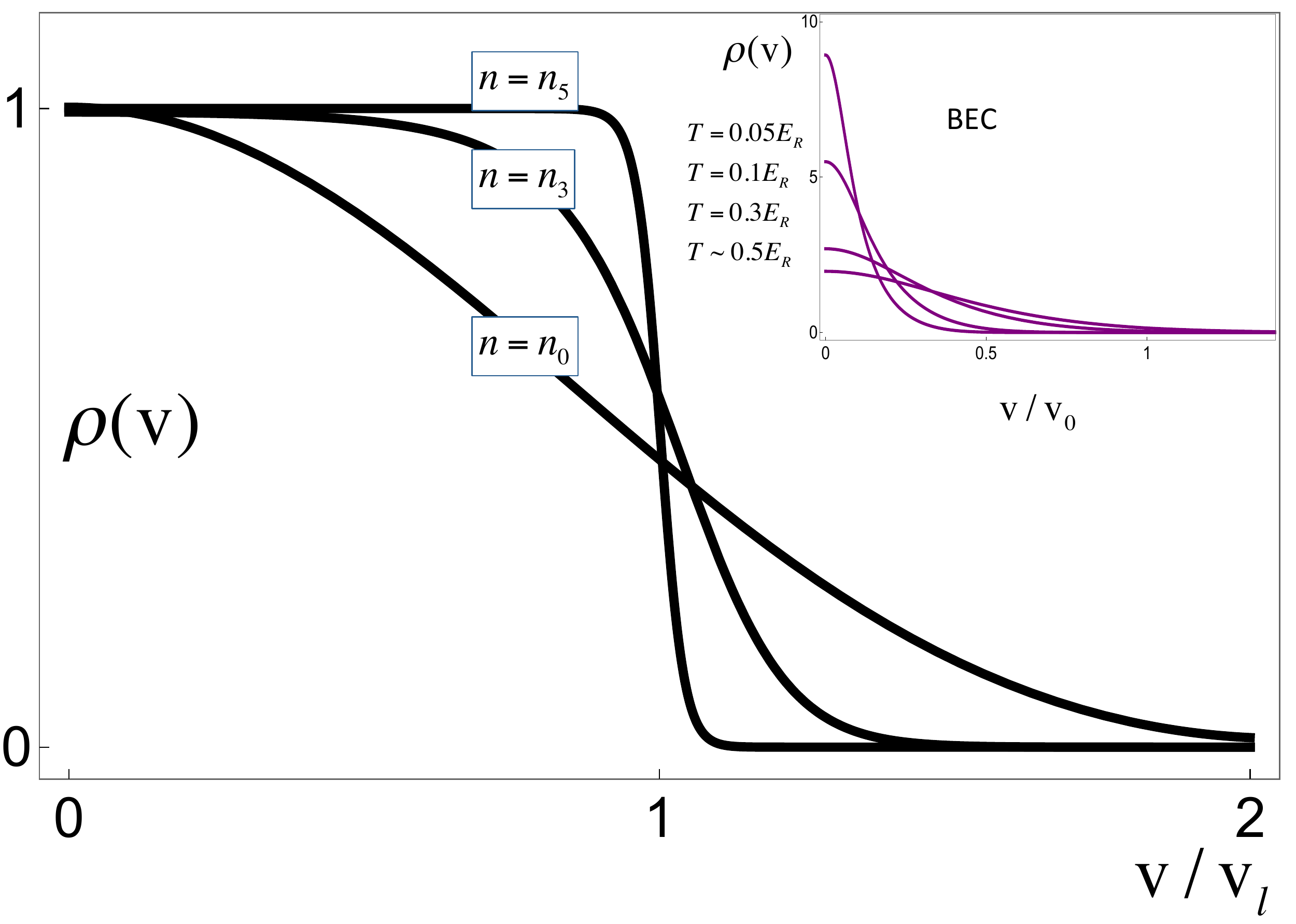}}
\caption{(Color online) Dimensionless velocity distribution $\rho(v)\cdot v_l$, where $mv_l=\pi n_l/k_0 $, of an expanding gas of composite fermions plotted vs dimensionless velocity $v/v_l$
at fixed values of density  $n=n_l$, $l=0,3,5$ and zero temperature. Inset: Dimensionless velocity distribution of condensed bosons with density
$n=n_0$. 
The temperatures are marked. At low temperatures the distribution shows a sharp peak at zero velocity, which is absent in the distribution of the composite fermions. }
\label{velo}
\end{figure}

An important experimentally relevant measure of the composite fermion state of bosons discussed above is the velocity 
distribution of an expanding gas, which can be observed in the time of flight experiments\cite{bloch}. 
 The group velocity of an expanding gas is defined by the derivative of the kinetic energy as
${\bf v}=\partial_{\bf k}\frac{(|{\bf k}|-k_0)^2}{2M}=\frac{|{\bf k}|-k_0}{M} \frac{\bf k}{|\bf k|}$. Expectation value of this operator in the proposed 
state of composite fermions (\ref{ca0}), (\ref{LL}) is obtained numerically and depicted in Fig.~\ref{velo}. 
The result 
demonstrates striking difference with the velocity distribution of condensed bosons shown in the inset of Fig.~\ref{velo}. 
While at high temperatures distribution functions of condensate and of composite fermion state are similar, 
the qualitative difference at $T\rightarrow 0$ is caused by the fermionic nature of the latter. If for condensed bosons the distribution is sharply peaked at 
$v=0$, indicating condensation into a state with zero velocity, for composite fermions it is reminiscent to the Fermi-Dirac 
distribution exhibiting weak, plateau-like behavior at finite $v$ at very low temperatures and small densities. Importantly, at low temperatures, 
there is no sharp peak at $v=0$. The plateau vs peak difference 
can be regarded as the indication of the proposed statistical transmutation.  
In the field-theoretical language this difference can be traced back to the presence of the effective Chern-Simons magnetic field and to the fact 
that effectively fermions find themselves in a state corresponding to the fully occupied lowest Landau level.

To conclude, we note that the ability to control and probe statistical transmutation in quantum
many body systems is one of the most fundamental challenges in contemporary physics. In this letter 
we propose an experimental scheme to {\em (i)} engineer 
a resonantly driven bosonic system exhibiting a moat-band and the phenomenon of transmutation of statistics, and {\em (ii)} probe our prediction for 
the velocity distribution in time of flight experiments
at low densities. The proposed state for bosons in a moat band is energetically more efficient 
at low densities than any other known candidate for the ground state. It
realizes a Floquet topological phase of bosons,  joining the family of other topological structures achieved in non-equilibrium including 
the  Floquet topological insulators\cite{lindner,podol,titum} and superfluids\cite{foster}.


This work was supported by the PFC-JQI (T.S.), USARO, Australian Research Council,
and Simons Foundation (V.G.), and DOE contract DE-FG02-08ER46482 (A.K.).


\pagebreak
\vspace{5cm}
\widetext
\begin{center}
\textbf{\large Supplementary material for: Statistical transmutation in Floquet  driven optical lattices}
\end{center}
\setcounter{equation}{0}
\setcounter{figure}{0}
\setcounter{table}{0}
\setcounter{page}{1}
\makeatletter
\renewcommand{\theequation}{S\arabic{equation}}
\renewcommand{\thefigure}{S\arabic{figure}}
\renewcommand{\bibnumfmt}[1]{[S#1]}
\renewcommand{\citenumfont}[1]{S#1}

\author{Tigran Sedrakyan}
\affiliation{William I. Fine Theoretical Physics Institute,  University of Minnesota, Minneapolis, Minnesota 55455, USA}
\affiliation{Physics Frontier Center and Joint Quantum Institute, University of Maryland, College Park, Maryland 20742, USA}

\author{Victor Galitski}

\affiliation{ Joint Quantum Institute and Condensed Matter Theory Center, Department of Physics,
University of Maryland, College Park, Maryland 20742-4111, USA}

\affiliation{School of Physics, Monash University, Melbourne, Victoria 3800, Australia}

\author{Alex Kamenev}
\affiliation{William I. Fine Theoretical Physics Institute,  University of Minnesota, Minneapolis, Minnesota 55455, USA}

\date{\today}

\maketitle

Here we address the energy $E$ of the proposed variational state and how does it scale with the particle density $n$, beyond the mean field approximation. 
To address these questions we develop a systematic approach to the calculation of the fluctuation effects.
While the mean-field predicts zero  energy for the  set of densities $n_l=k_0^2/\left[4\pi(l+1/2)\right]$, 
here we  show that the actual energy per particle $E = \langle (|k|-k_0)^2\rangle/2M$ calculated as the expectation value of the variational state (1)-(2) scales with density as $\sim n^2 \log^2 n$. At small density this energy is lower than that of both the bare condensate $\sim n$ and condensate renormalized by Cooper channel fluctuations\cite{Lam}, $\sim n^{4/3}$.


The possibility to analytically account for fluctuations comes from the fact that at discrete densities $n_l$, 
the mean field yields a state with zero-energy (calculated from the bottom of the band). The many-body bosonic wave function 
is given by 
\bea
\label{LL-1}
\Phi({\bf r}_1,\cdots {\bf r}_N)&=&\prod_{i<j}^N\frac{z_i-z_j}{|z_i-z_j|}\Psi_F({\bf r}_1,\cdots {\bf r}_N)
\ena
where $z=x+i y$ is the complex representation of the  2D vector  ${\bf r}=(x,y)$, and $l$ numerates the fully filled Landau level (LL). The wave function $\Psi_F({\bf r}_1,\cdots {\bf r}_N) $ is a fermion Slater determinant of spinless noninteracting fermions, residing on the $l$-th LL:
\bea
\label{LL-2}
\Psi_F({\bf r}_1,\cdots {\bf r}_N)&=&
\frac{1}{\sqrt{N!}} \mbox{det}_{m,j} \Big[\chi_{m}^{(l)}(z_j)\Big]
\ena
Here $\chi_{m}^{(l)}(z)$ is the single particle wave function corresponding to a state with angular momentum $m$. It is given by 
\bea
\label{LL-3}
\chi_{m}^{(l)}(z)=\frac{(-1)^l \sqrt{l!}}{l_B \sqrt{2 \pi 2^m (l+m)!}}\Big(\frac{z}{l_B}\Big)^m e^{\frac{|z|^2}{4 l_B^2}} 
L_{l}^{(m)}\Big[\frac{|z|^2}{2 l_B^2}\Big],\;\;\quad m=-l \cdots N-l.
\ena
where $L_{l}^{(m)}(x) $ is the adjoint Laguerre polynomial and $l_B=1/\sqrt{2 \pi n}$ is the magnetic length.

{\em Non degenerate spectrum:} As a first step we investigate the energy in the absence of the moat, $k_0=0$. It is given by the expectation value of the kinetic term $\langle{\bf k}^2\rangle $ (due to the antisymmetric fermionic multiplier of the wave function, the expectation value of the contact interaction potential is zero). In this expression, the Chern-Simons factor in (\ref{LL-1}) creates a $U(1)$ gauge vector
potential, $A_\mu({\bf r})=\epsilon_{\mu \nu} \frac{(r)_\nu}{|{\bf r}|^2} $. One observes that, 
 only  one, two, and
three particle states contribute to this expectation value. There are two types of terms contributing into 
$\langle{\bf k}^2\rangle =\langle {\bf k}^2\rangle_{diag}+\langle {\bf k}^2\rangle_{non-diag} $:
contributions coming from  diagonal in $m,{\bf r}$ variables states and non-diagonal ones. 
The general analytical expression for diagonal terms reads
\bea
\label{S-5}
\langle {\bf k}^2\rangle_{diag} = \frac{1}{N}\sum_{m}\int d{\bf r}\,\, \chi^*_{m}({\bf r}) \Big\{\big({\bf k}-{\bf {\cal A}}({\bf r})\big)^2 
+{\cal H}_{diag}({\bf r})\Big\} \chi_{ m}({\bf r})
\ena
where hereafter we suppress index $l$ in most intermediate expressions and the gauge field is given by
\bea
\label{S-6}
{\cal A}_\mu({\bf r})=\epsilon_{\mu \nu}\sum_{{\bf r'},m=-l}^{N-l}\langle m,{\bf r'}| A_\mu({\bf r}-{\bf r'}) |m,{\bf r'}\rangle  =\epsilon_{\mu \nu} \sum_{m=-l}^{N-l}\int d{\bf r'}\,
\chi^*_{m}({\bf r'})\frac{(r-r')_\nu}{|{\bf r}-{\bf r'}|^2} \,  \chi_{m}({\bf r'}). 
\ena
The function ${\cal H}_{diag}({\bf r})$ is given by
\bea
\label{S-7}
{\cal H}_{diag}({\bf r})&=&\sum_{m_1}\int d{\bf r}_1 |\chi_{m_1}({\bf r}_1)|^2{\bf A}^2({\bf r}-{\bf r}_1)  \nn\\
&-&  \sum_{m_1 ,m_2}\sum_{{\bf r}_1,{\bf r}_2}\Big\{\langle m_1,{\bf r}_1|\langle m_2,{\bf r}_2| {\bf A}({\bf r}-{\bf r}_1)  {\bf A}({\bf r}-{\bf r}_2)|m_2,{\bf r}_1\rangle  |m_1,{\bf r}_2\rangle \\
&+&\langle\{m_1,{\bf r}_1,m_2,{\bf r}_2\}|{\bf A}({\bf r}_1-{\bf r})
{\bf A}({\bf r}_1-{\bf r}_2)
+{\bf A}({\bf r}_2-{\bf r}_1)  {\bf A}({\bf r}_2-{\bf r})\Big]|\{m_1,{\bf r}_1, m_2,{\bf r}_1\}\rangle  \Big\},\nn
\ena
 where the bra (ket) states $\langle m,{\bf r}| $ ($|m,{\bf r}\rangle   $) represent the single particle function $\chi_{m}({\bf r}) $ ($\chi^*_{m}({\bf r}) $), while the two-particle state 
 $|\{m_1,{\bf r}_1,m_2,{\bf r}_2\}\rangle  $ is given by a 
Slater determinant of 
two wave functions $\chi_{m_1}({\bf r}_1)$ and $\chi_{m_2}({\bf r}_2)$  with the same LL index $l$.

It is important to notice that in the thermodynamic limit the vector potential ${\cal A}_\mu({\bf r})$ gives raise to a constant Chern-Simons magnetic field  
\bea
{\cal A}_\mu({\bf r})\vert_{N\rightarrow\infty}\rightarrow \frac{1}{2}\epsilon_{\mu\nu}{\bf r}_{\nu}B,
\quad B=2 \pi n. 
\ena
Although the expression Eq.~(\ref{S-7}) does not exhaust all contributions, the contribution of non-diagonal terms vanishes in the thermodynamic limit.  Thus we conclude that the first term in Eq.~(\ref{S-5}) exactly reproduces the mean-field result. 

\begin{figure}[t]
\centerline{\includegraphics[width=85mm,angle=0,clip]{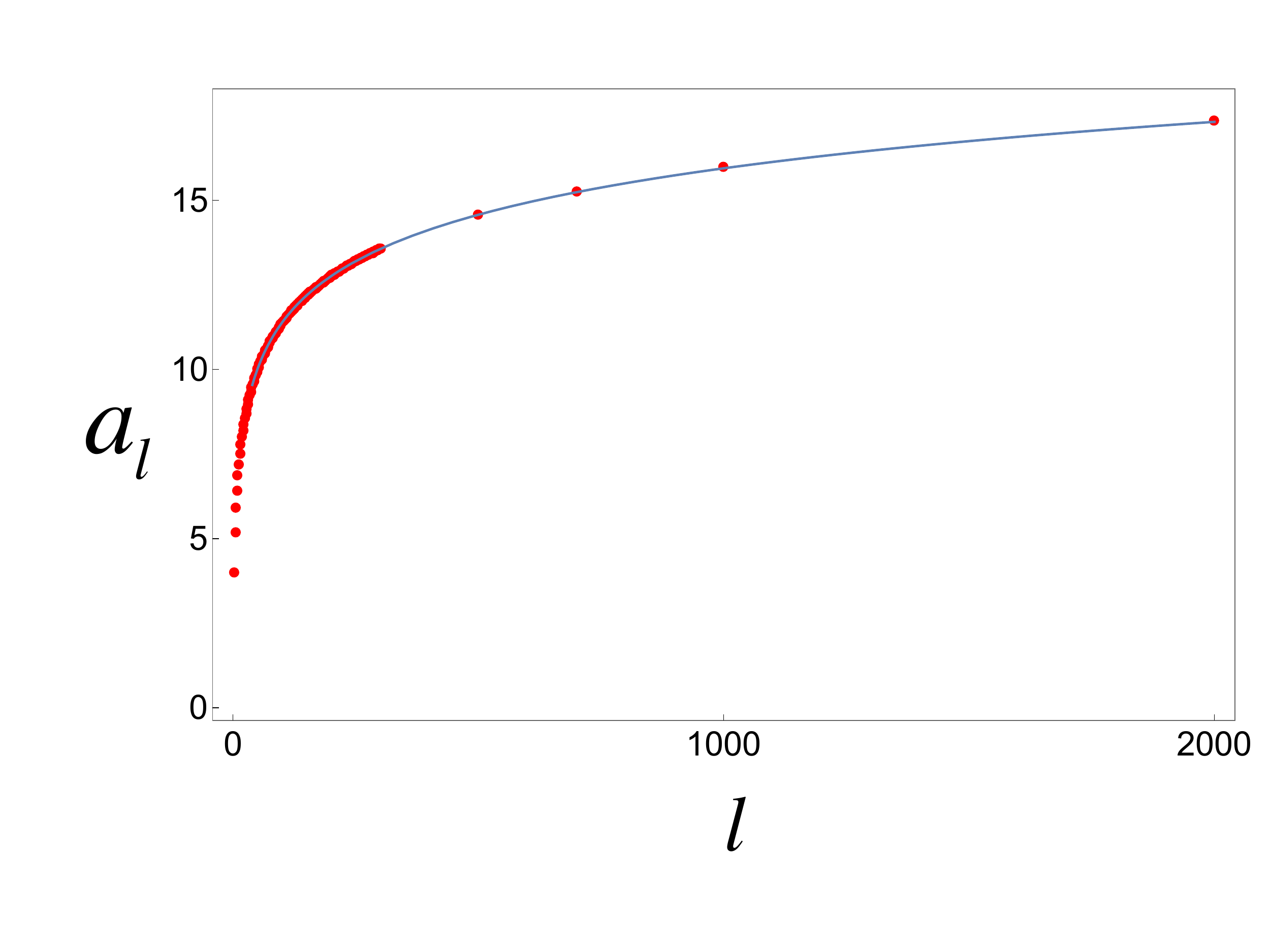}}
\caption{(Color online) Dots represent the coefficient $a_l$ plotted vs the number $l$ of the 
filled Landau level. 
The full line is the large $l$ asymptote which is consistent with $ a_l \approx 2 \log(\pi l)$. }
\label{H0}
\end{figure}

 The expression for the remaining non-diagonal  matrix elements of $\langle{\bf k}^2\rangle  $ can be cast as follows
\bea
\label{S-4}
&&\langle{\bf k}^2\rangle  _{non-diag}=- \frac{1}{N}\sum_{m,{\bf r}, m_1,{\bf r}_1}\langle m,{\bf r}|\langle m_1,{\bf r}_1|{\bf \big\{}{\bf k}, {\bf A}({\bf r}-{\bf r}_1){\bf \big\}}+{\bf A}^2({\bf r}-{\bf r}_1)
 |m_1,{\bf r}\rangle  | m,{\bf r}_1\rangle  \qquad \nn\\
&-&\frac{2}{N}\sum_{{\bf r},{\bf r}_1,{\bf r}_2} \sum_{m, m_1 ,m_2}\Bigl[ \langle m,{\bf r}|\langle m_1,{\bf r}_1|\langle m_2,{\bf r}_2| \Bigl[{\bf A}({\bf r}-{\bf r}_1)  {\bf A}({\bf r}-{\bf r}_2)+{\bf A}({\bf r}_1-{\bf r})
{\bf A}({\bf r}_1-{\bf r}_2)\qquad \\
&+&{\bf A}({\bf r}_2-{\bf r}_1)  {\bf A}({\bf r}_2-{\bf r})\Bigr]| m_1,{\bf r}\rangle  |m_2,{\bf r}_1\rangle  | m,{\bf r}_2\rangle  \Bigr],\nn
\ena
Where we used a notation ${\bf \big\{}{\bf k}, {\bf A}{\bf \big\}}={\bf k}{\bf A}+{\bf A}{\bf k}$.
Due to the orthogonality of the single particle wave functions $\chi_{m}^{(l)}({\bf r})$, most of this these non-diagonal terms 
are vanishing as $N\rightarrow \infty$.  Up to the terms vanishing in the thermodynamic limit $N\to \infty$,   the non-diagonal terms yield
\bea
\label{Snon-1}
&&\langle{\bf k}^2\rangle  _{non-diag}=- \frac{1}{N}\sum_{m,{\bf r}, m_1,{\bf r}_1}\langle m,{\bf r}|\langle m_1,{\bf r}_1|
{\bf A}^2({\bf r}-{\bf r}_1)|m_1,{\bf r}\rangle  | m,{\bf r}_1\rangle\nn\\
&=&-\frac{1}{N}\sum_m \int d{\bf r}|\chi_{m}({\bf r})|^2\int d{\bf r}_1 \frac{1}{2\pi l_B^2}{\bf A}^2({\bf r}-{\bf r}_1) 
e^{-\frac{|{\bf r}-{\bf r}_1|^2}{2 l_B^2}} L^2_l\Big[\frac{|{\bf r}-{\bf r}_1|^2}{2 l_B^2}\Big],
\ena
where we have used the completeness relation of the wave functions $\chi_{m}^{(l)}({\bf r})$\cite{LL,AG} 
\bea
\label{CR} 
\sum_m \chi_{m}^{(l)}({\bf r}_1)\chi_{m}^{*(l)}({\bf r}_2)=\frac{1}{2 \pi l_B^2}
e^{\frac{i}{2 l_B^2}(y_1-y_2)(x_1+x_2)-\frac{|{\bf r}_1-{\bf r}_2|^2}{4 l_B^2}}L_l \left[ \frac{|{\bf r}_1-{\bf r}_2|^2}{2 l_B^2}\right].
\ena
We see that, using the completeness relation, one can cast the non-diagonal term (\ref{Snon-1}) in 
the form similar to the diagonal one, as in Eq.~(\ref{S-5}).  
Joining Eq.~(\ref{Snon-1}) with ${\cal H}_{diag}({\bf r}) $ one obtains in the thermodynamic limit 
\bea
\label{Snon-2}
\langle {\bf k}^2\rangle = \frac{1}{N}\sum_{m}\int d{\bf r} \,\, \chi^*_{m}({\bf r})\, \Big[\big({\bf k}-{\bf {\cal A}}({\bf r})\big)^2 
+{\cal H}_{0}({\bf r})\Big] \chi_{m}({\bf r}),
\ena
where
\bea
\label{S-8}
{\cal H}_0({\bf r})&=& \int d{\bf r}_1 \frac{1}{2 \pi l_B^2}\Bigg[{\bf A}({\bf r}-{\bf r}_1)^2
\Big(1-e^{-\frac{|{\bf r}-{\bf r}_1|^2}{2 l_B^2}} L^2_l\Big[\frac{|{\bf r}-{\bf r}_1|^2}{2 l_B^2}\Big]\Big)
-\frac{1}{({\bf r}-{\bf r}_1)^2+2 l_B^2} \Bigg]\nn\\
&-& \int d{\bf r}_1 d{\bf r}_2 \frac{1}{(2 \pi)^2 l_B^4} e^{-\frac{|{\bf r}_1-{\bf r}_2|^2}{2 l_B^2}}
L_l^2\left[\frac{|{\bf r}_1-{\bf r}_2|^2}{2 l_B^2}\right] 
\Bigg({\bf A}({\bf r}-{\bf r}_1)  {\bf A}({\bf r}-{\bf r}_2)-\frac{1}{({\bf r}-{\bf r}_1)^2+2 l_B^2}\Bigg)\nn\\
&=&\frac{a_l}{2 l_B^2}.
\ena
Here we took into account the fact that last two terms in (\ref{S-7}) give vanishing contribution due to the antisymmetric nature of the Slater 
determinant. 
We also have added and subtracted $\frac{1}{({\bf r}-{\bf r}_1)^2+2 l_B^2} $ terms in both, first and second integrals which
explicitly shows convergence at large distances in each of them. Easy to see that the sum of the two added terms is identically equal to zero. 
The coefficient $a_l$ in Eq.~(\ref{S-8}) is  position independent as its ${\bf r}$-dependence may be absorbed into the shift of integration variables ${\bf r}_1$ and ${\bf r}_2$. 
The coefficients $a_l$ can be calculated analytically by using the explicit form of Laguerre polynomials  
\bea
\label{La}
L_l(z)=\sum_{k=0}^l \frac{(-1)^k l! z^k}{(l-k)! k!^2}.
\ena
Substituting Eq.~(\ref{La}) into Eq.~(\ref{S-8}), one is able to integrate the sum term by term. 
This results in the following series representation for the coefficient $a_l$
\bea
\label{coef}
a_l = 1 + \sum_{k_1,k_2=0,k_1+k_2\neq 0}^l \frac{(-1)^{k_1+k_2} l!^2 (k_1+k_2)!}{(l-k_1)!(l-k_2)!k_1!^2 k_2!^2} H_{k_1+k_2-1},
\ena 
where $H_k=\sum_{j=1}^k \frac{1}{j}$ are the Harmonic numbers. Coefficient $a_l$ is depicted vs $l$ in Fig.~\ref{H0}.
With an excellent prescision  $a_l$ is fit with the function 
$$ a_l \approx 2 \log(\pi l).$$

{\em Energy in the presence of a moat-band:}
In contrast to the case with non-degenerate dispersion, a moat-band spectrum studied in this paper exhibits a singularity.
The $|k|$ singularity in the average kinetic energy $\langle \epsilon({\bf k})\rangle  =\langle (|{\bf k}|-k_0)^2\rangle $ 
is however treatable. It can be dealt with by representing as 
\bea
\label{S-11}
\sqrt{{\bf k}^2}= {\bf k}^2 \int_0^\infty ds e^{- {\bf k}^2 s^2}.
\ena 
Using such a representation, one can straightforwardly extend the expression (\ref{S-5}) to the case with a moat. One can see 
that the ${\bf k}$ dependence of an integrand in the rhs of Eq.~(\ref{S-11}) is free from singularities. Thus, using the procedure discussed in the previous section one obtains
\bea
\label{S-12}
\langle (|{\bf k}|-k_0)^2\rangle    = \frac{1}{N}\sum_{m}\int d{\bf r}\,\, \chi^*_{m}({\bf r}) \Big[\sqrt{\big({\bf k}-{\bf {\cal A}}({\bf r})\big)^2 
+{\cal H}_0({\bf r})}-k_0\Big]^2 \chi_{ m}({\bf r}).
\ena   
Assuming that the contribution to the energy coming from ${\cal H}_0({\bf r})$ is much smaller than $k_0^2$ (this assumption however needs to be checked self-consistently) and expanding the square root in (\ref{S-12}) over ${\cal H}_0({\bf r})/k_0^2$,  one obtains
\bea
\label{S-13}
\langle (|{\bf k}|-k_0)^2\rangle  = \frac{1}{N}\sum_{m}\int d{\bf r}\,\,  \chi^*_{m}({\bf r}) \Big[\sqrt{\big({\bf k}-{\bf {\cal A}}({\bf r})\big)^2}
-k_0 +\frac{{\cal H}_0({\bf r})}{2 \sqrt{\big({\bf k}-{\bf {\cal A}}({\bf r})\big)^2}}+\cdots \Big]^2 \chi_{m}({\bf r}).
\ena   
At densities $n_l=k_0^2/\left[4\pi(l+1/2)\right] $, when the mean-field  yields zero energy  measured from the bottom of the moat, 
one can separate the mean-field result from the fluctuation corrections. With the help of the identity 
$ \sqrt{\big({\bf k}-{\bf {\cal A}}({\bf r})\big)^2}|m,{\bf r}\rangle \equiv k_0|m,{\bf r}\rangle$ (which is the statement of the zero-energy within the mean-field) and   Eq.~(\ref{S-8}) the leading fluctuation correction is found to be  
\bea
\label{S-14}
E=\frac{1}{2M}\langle (|{\bf k}|-k_0)^2\rangle  
\!\!\!&&=\frac{1}{2MN}\sum_{m}\int d{\bf r}\, \chi^*_{m}({\bf r})\, \left(\frac{{\cal H}_0}{2 k_0}\right)^2 \,\chi_{m}({\bf r})\nn\\
&&\simeq \frac{1}{8M l_B^4} \left(\frac{a_l}{2 k_0}\right)^2 \simeq \frac{\pi ^2}{2} \left( \frac{n^2}{Mk_0^2}\right) \log ^2(4n/k_0^2) .
\ena   
Eq.~(\ref{S-14}) constitutes the main result of the Supplementary material.
The validity of self-consistent assumption allowing to neglect higher order terms 
in Eq.~(\ref{S-13}) is guaranteed when the scale $\sim (n/k_0^2) \log^2n$ is smaller than one. This implies that the fluctuation induced energy (\ref{S-14}) is smaller
than $\sim n/M$, which is satisfied precisely in the low density limit when the composite fermion state is more efficient than a condensate. 

If the number of particles is finite, the fluctuation effects give raise to $N$-dependent finite-size corrections to the energy. 
These corrections are vanishing in the thermodynamic limit $N\rightarrow \infty$. Our analysis shows that the latter 
however reaches slowly. There are corrections to the energy (\ref{S-14}) originating from $\langle{\bf k}^2\rangle _{non-diag}$ that scale as $\sim n/(MN)$.
There is also possibility of logarithmical corrections suggesting that in order to reach the thermodynamic limit numerically 
one has to take exponentially large systems, $N \gtrsim \exp\{k_0^2/n\}$. 


\end{document}